\documentstyle[aps,prl,multicol,epsfig]{revtex}
\begin{document}
\draft
\widetext
\title{ Weak pseudogap behavior  in the underdoped cuprate superconductors  }
\author{J\"org Schmalian,$^{a}$ David Pines,$^{a,b}$
 and   Branko Stojkovi\'c$^{b}$    }
\address{$^{a}$University of Illinois at 
Urbana-Champaign,   Loomis Laboratory 
of Physics, 1110 W. Green,  Urbana, IL, 61801\\
$^{b}$Center for Nonlinear Studies, Los Alamos National Laboratory, 
Los Alamos, NM, 87545 }
\date{\today}
\maketitle
\widetext 
\begin{abstract} 
\leftskip 54.8pt
\rightskip 54.8pt
We report on an  exact solution of the nearly antiferromagnetic Fermi liquid 
 spin fermion model in the limit $\pi T \gg \omega_{\rm sf}$,
which demonstrates that  the broad  high energy  features
found in ARPES measurements  of the  spectral density of the
underdoped cuprate superconductors are
determined by  strong  antiferromagnetic (AF) 
correlations and precursor effects
of an SDW state.  
We show that the onset temperature, $T^{\rm cr}$, of weak pseudo-gap 
(pseudoscaling) behavior is determined by the strength, $\xi$, of the 
AF correlations, and obtain the generic changes in low frequency magnetic 
behavior seen in NMR experiments with
 $ \xi(T^{\rm cr}) \approx 2$, confirming
the  Barzykin and Pines crossover criterion.
\end{abstract}   
\pacs{74.25.-q,75.25.Dw,74.25.Ha} 
\begin{multicols}{2}
\narrowtext   
Magnetically underdoped cuprates
may be distinguished from their overdoped counterparts 
by the presence of a maximum
at a temperature $T^{\rm cr} > T_c$ in the temperature dependent uniform
susceptibility, $\chi_o(T)$.
They  are characterized by the occurrence of a quasiparticle
pseudogap observed by
NMR and INS experiments, optical,   transport  and specific heat measurements,
and in angular resolved photoemission spectroscopy (ARPES).
Barzykin and Pines~\cite{BP95} proposed that at $T^{\rm cr}$, 
sizable AF correlations between the planar quasiparticles 
bring about a change in the 
spin dynamics, and that at $T$ near  
$T^{\rm cr}$, the quantity 
$^{63}T_1T/^{63}T_{\rm 2G}^2$, where $^{63}T_1$ is
the $^{63}$Cu spin-lattice relaxation time and 
$^{63}T_{\rm 2G}$ is 
the spin-echo decay time, changes from being nearly 
independent of temperature (above $T^{\rm cr}$),
to a quantity which varies as $(a+bT)^{-1}$.
This behavior has recently been confirmed in 
NMR measurements by Curro {\em et al.}~\cite{CCS96}.
Since  $^{63}T_1T/^{63}T_{\rm 2G}^n\propto \xi^{n-z}$, where
 $z$ is a dynamical exponent\cite{BP95}, 
the near temperature independence of 
 $^{63}T_1T/^{63}T_{\rm 2G}$ found between $T^{\rm cr}$ and a lower
crossover temperature, $T_*$, suggests that one is in a 
pseudoscaling regime ($z\approx 1$), while the temperature independence of 
of  $^{63}T_1T/^{63}T_{\rm 2G}^2$ above $T^{\rm cr}$ suggests mean field
behavior ($z \approx 2$).
Moreover, above $T_*$ ARPES experiments show that the spectral
density of quasiparticles located near $(\pi,0)$ has developed
a high energy feature~\cite{ZX95}.
We refer to this behavior as 
{\em weak pseudogap behavior}, to distinguish
 it from the {\em strong pseudogap behavior}
 found below  $T_*$, where 
 experiment shows a leading edge gap develops
in the quasiparticle spectrum~\cite{ZX95}.
 Strong pseudogap behavior is also
 seen in specific heat, d.c.\ transport, and optical experiments, 
while $^{63}T_{\rm 2G}$ measurements show that
the AF spin correlations become frozen (i.e., $\xi \approx const.$)
 and $^{63}T_1T$ displays  gap-like behavior.

The nearly antiferromagnetic Fermi liquid (NAFL)
 model~\cite{MBP93,MP93} of the cuprates offers a possible explanation
for the observed weak and strong pseudogap
behavior.
 In this model, changes in quasiparticle behavior
 both reflect and bring about the measured changes in spin dynamics.
The highly anisotropic
 effective planar quasiparticle interaction mirrors the
 dynamical spin susceptibility, 
peaked near ${\bf Q}=(\pi.\pi)$, introduced by Millis,
 Monien, and Pines~\cite{MMP90} to explain NMR experiments:
\begin{equation}
V_{\rm eff}^{\rm NAFL}({\bf q},\omega)=g^2 \chi_{\bf q}(\omega)=\frac{g^2
\chi_{\bf Q}}{1+\xi^2({\bf q}-{\bf Q})^2
- i {\omega\over \omega_{\rm sf}}}\, ,
\label{MMP}
\end{equation}
where   $\chi_{\bf Q}=\alpha \xi^2$, with $ \alpha $  constant,
and $g$ is the coupling constant.

Since the dynamical spin susceptibility $\chi_{\bf q}(\omega)$ peaks  
 at wave vectors
close to $(\pi,\pi)$, two different kinds of 
quasiparticles emerge: {\em hot quasiparticles}, located 
close to those  momentum points on the Fermi surface which
 can be  connected by ${\bf Q}$,
feel the full effects of the interaction of Eq.\ref{MMP};  {\em cold
quasiparticles}, located not far from  the diagonals, 
$\vert k_x\vert=\vert k_y\vert$, feel a ``normal''  interaction.
Their distinct lifetimes can be inferred from transport experiments,
where a detailed analysis shows that
the behavior of the hot quasiparticles is highly  anomalous, 
while cold quasiparticles 
 may be characterized as a strongly  coupled Landau Fermi Liquid~\cite{SP96}.

In the present communication we focus our
 attention on temperatures $T\geq T_*$.
Our reason for doing so is that for $T>T_*$, 
fits to NMR experiments show that $\omega_{\rm sf}< \pi T$, the frequency
equivalent of temperature.
Because of  its comparatively  
low characteristic energy  the spin system
 is thermally excited and behaves quasistatically;
the hot quasiparticles see a
spin system which  acts like a static  deformation potential, 
a  behavior 
which is no longer found below $T_*$  where the lowest
scale is the temperature itself and the quantum nature of the spin degrees of freedom is essential.
We find that above $T_*$,
in the limit $\pi T \gg \omega_{\rm sf}$ 
it is possible to obtain an
 exact solution of the spin fermion model with the  effective 
interaction,  $V_{\rm eff.}^{\rm NAFL}({\bf q},\omega)$,
of Eq.~\ref{MMP}.

Our  main results are the appearance 
in the hot quasiparticle spectrum of the high energy features seen in ARPES, 
a  maximum in $\chi_o(T)$ and a crossover 
in $^{63}T_1T/^{63}T_{\rm 2G}^2$ for
 $\xi > \xi_o \approx v_{\rm F}/ \Delta_o $. 
These are  produced by the emergence of an SDW-like state,
as proposed by Chubukov {\em et al.}~\cite{CPS96}
Here, $v_{\rm F}$ is the Fermi velocity and 
$\Delta_o=\frac{g}{\sqrt{3}} \sqrt{\langle {\bf S}^2 \rangle} 
\sim \frac{g}{2}$, a characteristic energy scale of the 
SDW-like pseudogap.
Using typical values for the hopping 
elements (see below), and $g \approx 0.6\, {\rm eV}$ (determined
from the analysis of transport experiments in slightly underdoped
materials~\cite{SP96}), we find $\xi_0 \approx 2$.
For $\xi >\xi_o$,  the hot quasiparticle  spectral density 
takes a  two-peak form which reflects  the emerging 
 spin density wave state, while 
the MMP interaction generates naturally the 
distinct behavior of hot and cold quasiparticle states
seen in  ARPES experiments.

Before discussing these results, we summarize our calculations briefly.
Using the effective interaction,   Eq.\ \ref{MMP},
the  direct spin-spin  coupling is eliminated via a
  Hubbard-Stratonovich transformation, introducing
a collective spin field ${\bf S}_{\bf q}(\tau)$. 
After integrating out the fermionic degrees of freedom,   the single
particle Green's function can be written as
\begin{equation}
G_{{\bf k},\sigma} (\tau - \tau') =
 \left\langle
 \hat{G}_{{\bf k},{\bf k} \sigma \sigma }(\tau,\tau' |{\bf S} )
 \right\rangle_o \, ,
\label{GF}
\end{equation}
where $\hat{G}_{{\bf k},{\bf k} \sigma \sigma }(\tau,\tau'
 |{\bf S} )$ is the matrix element of
 \begin{eqnarray}
[G^{-1}_{o{\bf k} }(\tau-\tau' ) \delta_{{\bf k},{\bf k'}} 
-\frac{g}{\sqrt{3}} {\bf S}_{{\bf k} -
{\bf k'}}(\tau)\delta(\tau-\tau')\cdot {\bf \vec\sigma}]^{-1} ,
\label{GFmatrix}
\end{eqnarray}
which describes the propagation of an electron for a given configuration ${\bf S} $ of the spin field.
${\bf \vec\sigma}$ is the Pauli matrix vector  and
 $G^{-1}_{o {\bf k}}=-(\partial_\tau +\varepsilon_{\bf k} )$
is the inverse  of the unperturbed  single particle Green's function with bare dispersion
\begin{equation}
\varepsilon_{\bf k}=-2t(\cos k_x +\cos k_y ) -4t' \cos k_x \cos k_y -\mu \, .
\end{equation} 
In the following we use $t=0.25\, {\rm eV}$ and $t'=-0.4 t$ 
for the nearest and next nearest
neighbor hopping integrals, respectively, and 
we adjust the chemical potential $\mu$ to maintain the
constant hole concentration at $n_h=15$\%.
The average 
$ \langle \cdots \rangle_o= \frac{1}{Z_{\rm B}} \int {\cal D}{\bf S} \cdots
\exp\left\{-S_o\right\} $
 is performed  with respect to the action  of the collective spin
 degrees of freedom:
\begin{equation}
S_o({\bf S})=\frac{T}{2} 
\sum_{{\bf q},n} \chi_{\bf q}^{-1}(i\omega_n) \,
 {\bf S}_{\bf q}(i\omega_n) \cdot {\bf S}_{-{\bf q}} (-i\omega_n) 
\label{coll0}
\end{equation}
where $\omega_n=2n\pi T$
and $Z_{\rm B}$ is   defined via $ \langle 1 \rangle_o=1$.
In using  Eq.~\ref{GF} we have assumed that  ({\em i})
$\chi_{\bf q}(\omega)$ is the fully renormalized spin-susceptibility
taken from the experiment and 
  ({\em ii})    any nonlinear (higher order in ${\bf S}$ than quadratic)  terms
of the spin field  can be neglected.
 The model which  results  from assumption  ({\em ii})  is usually referred to
as the spin fermion model.  

After inversion of Eq.~\ref{GFmatrix} in spin space,
the average of  Eq.~\ref{GF} can be  evaluated   diagrammatically
 using Wick's theorem for  the spin field.
In the above mentioned static limit, $\pi T \gg \omega_{\rm sf}$, it 
 suffices to
consider only the zeroth bosonic
 Matsubara frequency in $\chi_{\bf q}(i\omega_n)$.
The remaining momentum summations are evaluated  by expanding 
$\varepsilon_{{\bf k} + {\bf q}} \approx 
\varepsilon_{{\bf k} + {\bf Q}} + {\bf v}_{{\bf k} + {\bf Q}} \cdot
({\bf q}-{\bf Q})$ for momentum transfers close to ${\bf Q}$, using
$v^\alpha_{{\bf k} + {\bf Q}}=\partial \varepsilon_{{\bf k} + {\bf Q}}/
\partial k_\alpha$.
In this limit {\em all diagrams} can be summed up by 
generalizing    a  solution  for  a one dimensional
charge density wave system obtained  by Sadovskii~\cite{Sad79} to 
the case of two dimensions, and more
 importantly, to isotropic spin fluctuations.
We find the following  recursion relation for  the  Green's function
 $G_{\bf k}(\omega) \equiv G^{l=0}_{\bf k}(\omega)$,
whose imaginary part determines the spectral density $A({\bf k},\omega)$,
 seen in ARPES\cite{comment}:  
\begin{equation}
[{G^l_{\bf k}(\omega)}]^{-1}={\omega -
\varepsilon_{{\bf k}+l{\bf Q}}+i{l v_{{\bf k},l}\over \xi} -
\kappa_{l+1}\Delta_o^2 G^{l+1}_{\bf k}(\omega)}.
\label{sadself}
\end{equation}
Here, $v_{{\bf k},l}=|{\bf v}_{{\bf k} 
+ {\bf Q}} |$  and $\kappa_l=(l+2)/3$ 
if $l$ is odd, while
$v_{{\bf k},l}=|{\bf v}_{{\bf k} + 
{\bf Q}} |(| {\rm cos}\phi_{\bf k} |+ |{\rm sin}\phi_{\bf k} |)$
and $\kappa_l=l/3$ if $l$ is even. 
 $\phi_{\bf k} $ is the angle between 
${\bf v}_{{\bf k}+{\bf Q}}$ and ${\bf v}_{{\bf k} }$.
The recursion relation, Eq.\ (\ref{sadself}), enables us 
to calculate $A({\bf k},\omega)$ to arbitrary order in
the coupling constant $g$. In the limit $\xi\rightarrow \infty$
the Green's function reduces to 
$G_{\bf k}(\omega) = \int d\Delta\ p(\Delta)\ 
G^{{\rm SDW}}_{\bf k}(\Delta)$, where $G^{{\rm SDW}}_{\bf k}(\Delta)$ is
the single particle Green's function of the mean field SDW state
and $p(\Delta)\sim\Delta^2 \exp(-{3\over 2} \Delta^2/\Delta_0^2)$
is the distribution function of a fluctuating SDW gap,
centered around $\sqrt{{2\over 3}}\Delta_0$,
i.e., the amplitude 
fluctuations of the spins ${\bf S}$ are  confined to a region  around 
$\sqrt{\frac{2}{3} \langle{\bf S}^2\rangle }$, although 
in our calculations
directional fluctuations are fully isotropic and spin rotation
invariance is maintained.
Below we show that the SDW like solution is
obtained even at finite values of $\xi$.

The quantities we calculate are the single particle spectral 
density, $A({\bf k},\omega)$ and the low frequency behavior of
the  irreducible spin
 susceptibility
$\tilde{\chi}_{\bf q}(\omega, T)$.
Our principal results are depicted in Figures 1-3.
The $\xi$-dependence of the Fermi surface, shown in the inset to Fig.\ 1a, 
is similar to the results obtained by Chubukov
{\em et al.}~\cite{CM97} at $T=0$; 
however for the experimentally 
relevant range of $\xi$, hole pockets around 
${\bf k}=(\pi/2,\pi/2)$ do not form. We have used the criterion
$\varepsilon_{\bf k}+{\rm Re}\Sigma_{\bf k}(\omega=0)=0$ to
determine  the Fermi surface. Although strictly valid only for $T=0$, 
for finite temperatures it indicates when  a 
quasiparticle crosses the chemical potential.
In the limit of very large correlation length, we find, 
in addition to  the two   broadened  poles of a SDW like state, 
a third solution of 
$\omega={\rm Re} \Sigma_{\bf k}(\omega)+\varepsilon_{\bf k}$, which
although  not visible due to the large scattering rates,  
ensures that  even for $\xi \rightarrow \infty$ a large Fermi surface and 
only a pseudogap in  the density of states occur.

A comparatively sharp transition between the 
behavior of hot quasiparticles 
(located at points $a$ and $b$ on the Fermi surface of
Fig.\ 1a) and cold quasiparticles (at $d$ and $e$) is found. 
For hot quasiparticles the single particle spectral 
density evolves with temperature as
 the AF correlation length increases from 
$\xi \approx 1$ to $5$; a two peak structure, 
which corresponds to a transfer of spectral weight from 
low frequencies to frequencies 
above $200 - 300 \, {\rm meV}$, develops at $\xi_o \approx 2$, 
and is quite pronounced for $\xi \ge 3$ (Fig.\ 1b).
As may be seen in Fig.\ 1a, the shift in spectral
 density found for hot quasiparticles does not occur for 
cold quasiparticles, whose spectral density
 continues to be peaked at the Fermi energy.
\begin{figure}
\centerline{\epsfig{file=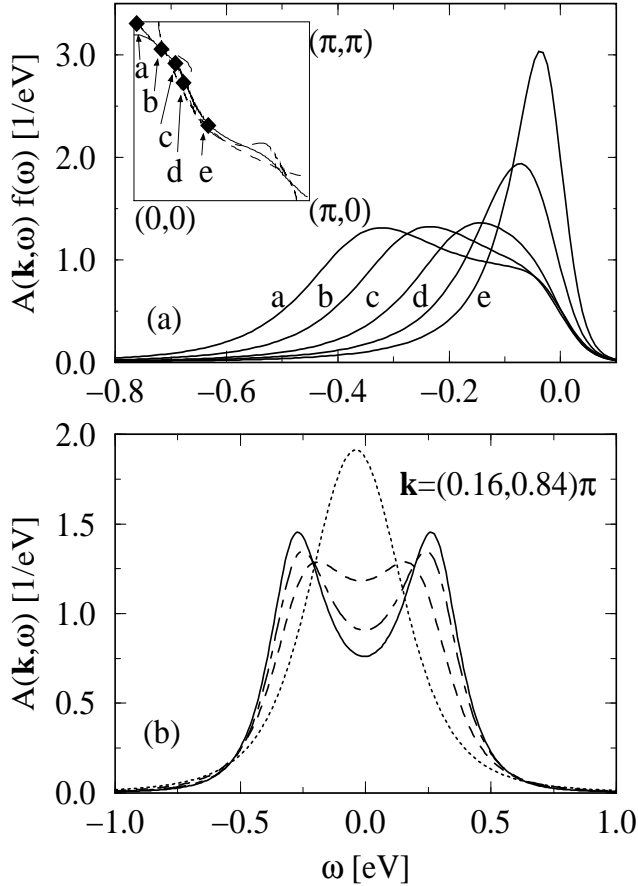,width=8.5cm, height=12cm, scale=0.85}}
 \caption{(a) Spectral density multiplied with Fermi function on the 
 Fermi surface  for  $\xi=3$. 
The distinct behavior of  hot and cold quasiparticles is visible.
The inset shows the corresponding Fermi surface and its
 $\xi$-dependence. The dashed, solid and dashed-dotted lines correspond to
$\xi=1$, 3 and 10, respectively.
(b) $\xi$-dependence  of the spectral density of point $b$ in (a).
 The transition from a conventional line shape for
small $\xi=1$ (dotted line) to spectral densities with broad 
lines and pronounced high energy features ($\xi=2,3$, and $5$ for the
dashed, dashed dotted, and solid line, respectively),
 reflecting a precursor SDW state, is visible.
 Note, the saturation of the precursor effects for $\xi \approx 3$.}
\label{fig1}
\end{figure}

We sum {\em all diagrams} of
the  perturbation series for the electron-spin fluctuation 
vertex function in similar fashion as the Green's function $G_{\bf k}(\omega)$
in Eq.\ (\ref{sadself}).
The lack of symmetry breaking is essential for a
proper evaluation of the vertex which,
as long as the spin rotation invariance is intact, is
reduced at most  by a factor $\approx \frac{1}{3}$
for the high energy features\cite{CM97,Bob}.
For lower excitation energies, this vertex is  considerably enhanced
for the hot quasiparticles; it is  almost unaffected for the cold
quasiparticles, reflecting again their qualitatively different behavior.

We combine the results for $G_{\bf k}(\omega)$
and the electron-spin fluctuation vertex function
and so determine the irreducible spin susceptibility
$\tilde{\chi}_{\bf q}(\omega)$. 
We find (Fig.\ 2)  that both
 $\tilde{\chi}_o(T)$ and $\tilde{\chi}_{\bf Q}(T)$ 
exhibit maxima at temperatures close to $T^{\rm cr}$ where
 $\xi \approx 2$.
In these calculations we assumed 
that $\xi^{-1}(T)=\frac{1}{4}+\frac{1}{4}\frac{T-T_*}{T^{\rm cr}-T_*}$
between $T^{\rm cr}=470 \, {\rm K}$
and $T_* =220 \, {\rm K}$ and
$\xi^{-2}(T)=\frac{1}{4}+\frac{1}{7} 
\frac{T-T^{\rm cr}}{700\, {\rm K}-T^{\rm cr}}$
above $T^{\rm cr}$ consistent with the NMR 
results of Curro {\em et al.}~\cite{CCS96} for
YBa$_2$Cu$_4$O$_8$. 
The behavior of   $\tilde{\chi}_o(T)$ and $\tilde{\chi}_{\bf Q}(T)$
 above the maximum reflects the increasing importance of lifetime
 (strong coupling) effects which act
 to reduce both irreducible susceptibilities.
Because of comparatively short correlation lengths ($\xi <2$), it
is likely that Eliashberg calculations~\cite{MP93}
 will provide a better quantitative account
 in this mean field  regime. The fall-off
in  $\tilde{\chi}_o(T)$ and $\tilde{\chi}_{\bf Q}(T)$ below $T^{\rm cr}$
arises primarily from
the transfer of quasiparticle spectral weight
 to higher energies.

The determination of the
 full spin susceptibility 
$\chi_{\bf q}(\omega)=\tilde{\chi}_{\bf q}(\omega)/
(1-J_{\bf q}\tilde{\chi}_{\bf q}(\omega))$  requires calculating 
 the restoring force
$J_{\bf q}$, and is beyond the scope of the present work,
 since $J_{\bf q}$ is 
determined by the renormalization of the spin exchange fermion-fermion
interaction through high energy excitations in all other channels. 
However, for $\chi_o(T)$, with the assumption that 
$J_{{\bf q}=0}\tilde{\chi}_0(T_{cr})=0.5$,
a good quantitative fit to the experimental results of
Curro {\em et al.}~\cite{CCS96} 
between $T^{\rm cr}$ and $T_*$ is found.
\begin{figure}
\centerline{\epsfig{file=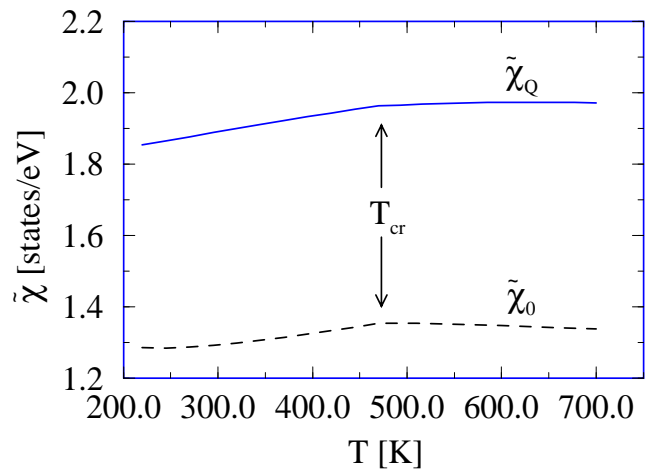,width=8.5cm, height=6.5cm, scale=0.85}}
\caption{Irreducible part $\tilde{\chi}_{\bf q}(T)$ of the static 
spin susceptibility 
  at ${\bf q}=(0,0)$ and ${\bf q}=(\pi,\pi) $ as function of temperature. }
\label{fig2}
\end{figure}
We find that both vertex corrections and quasiparticle spectral
weight transfer play a significant role in determining the low
frequency spin dynamics. 
As may be seen in the inset to Fig.\ 3, 
when both effects are taken into account,
our calculated values of the spin damping 
$\gamma_{\bf Q}=\left.
 \tilde{\chi}''_{\bf Q}(\omega,T)/\omega \right|_{\omega=0}$
display the crossover at $T^{\rm cr}$ anticipated by Monthoux and
 Pines~\cite{MP94}. 
A second calculable quantity which can be compared with experiment is 
$\tilde{\chi}_{\bf Q}(T)^2/\gamma_{\bf Q} \equiv \omega_{sf} \chi_{\bf Q}$,
being proportional to the product, $^{63}T_1 T/(^{63}T_{\rm 2G})^2$. 
As may seen in Fig. 3, qualitative agreement with the results of 
Curro {\em et al.}~\cite{CCS96} for YBa$_2$Cu$_4$O$_8$ is found.

Physically, the most interesting aspect of 
our results is
 the appearance of SDW  precursor phenomena,
brought about 
 by the strong interaction between the planar quasiparticles,
 for  moderate AF correlation lengths, $\xi > \xi_o \approx 2$,
in contrast to earlier calculations, in which 
SDW precursor behavior was only found in the limit of 
very large correlation length~\cite{KS90}.
Our exact solution of the static problem enables us to access this region
of strong coupling. For $\xi > \xi_o$, the 
hot electron mean free path,  $\sim \xi_o^2/\xi$
begins to be   small compared to $\xi$, so that
 the quasi particle can no longer  distinguish the
 actual situation from that of a
SDW state; hence we find  pseudo-SDW behavior, i.e. SDW behavior 
without symmetry breaking.
The related shift of spectral weight for states close to $(\pi,0)$
affects mostly the low frequency part of the irreducible spin susceptibility
and leads to the calculated  crossover behavior.
Because  symmetry is not broken, there will continue to be
coherent states  at the Fermi energy; 
although their spectral weight is small one still has a large Fermi surface.
Such  coherent quasiparticles, invisible in the present temperature range,
are, we believe,  responsible for the sharp peak for ${\bf k}\sim (\pi,0)$,
observed in ARPES below 
the superconducting transition temperature\cite{ZX95}.
\begin{figure}
\centerline{\psfig{file=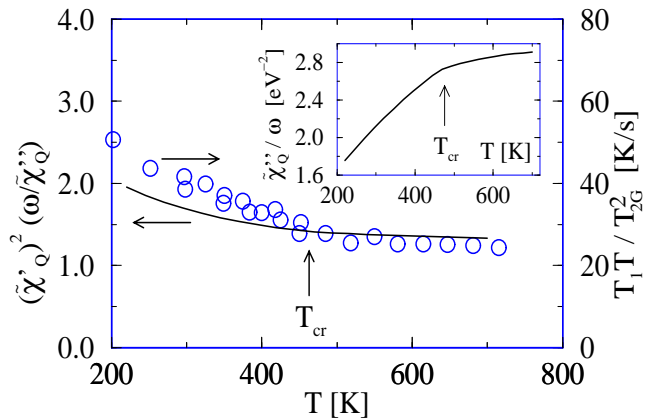,width=8.5cm, height=6.1cm, scale=0.85}}
\caption{
 $\left .\tilde{\chi}_{\bf Q}^2
 \omega /\tilde{\chi}''_{\bf Q} \right|_{\omega=0}$ 
as function of temperature, compared with experimental results of Ref.[2] for 
$T_1T/T_{\rm 2G}^2$.
The inset shows the crossover in the calculated $T$-dependence of 
$\left. \tilde{\chi}''_{\bf Q}(\omega)/\omega \right|_{\omega=0}$.} 
\label{fig3}
\end{figure} 

The present theory cannot, of course, explain the leading edge pseudogap
found below $T_*$,
since one is then no longer in the quasistatic 
limit for which our calculations apply.
Strong pseudogap behavior corresponds to a further redistribution
of quasiparticle states lying within $\approx 30 \, {\rm meV}$
of the Fermi energy.
No appreciable change is seen in the high energy features
found in the present calculations.
It is likely that strong scattering in the particle-particle channel
 plays an increasingly important role below $T_*$,
since we find above $T_*$ important prerequisites for its appearance;
 an enhanced  spin fluctuation vertex and a  pronounced flattening 
of the low energy part of the  hot quasiparticle  states~\cite{Cub97}.
In addition,  below $T_*$ the quantum
behavior of spin excitations becomes increasingly 
important. This reduces
the phase space for quasiparticle scattering, and leads to the
 sizable suppression of the hot quasiparticle scattering rate found
below  $T_*$~\cite{SP96}.

This work has been supported in part by the Science and Technology
Center for Superconductivity through NSF-grant DMR91-20000, 
by the Center for Nonlinear Studies at Los Alamos National Laboratory,
and by the Deutsche Forschungsgemeinschaft (J.S.).
We thank the Aspen Center for Physics for its hospitality
during the period in which part of this paper has been written,
and Andrey Chubukov for helpful discussions.
  

\end{multicols}

\begin{thebibliography}{99}
\bibitem{BP95} V. Barzykin and D. Pines, Phys. Rev. B {\bf 52}, 13585 (1995).
\bibitem{CCS96}N. J.  Curro, T. Imai, C. P. Slichter,
and B. Dabrowski, Phys. Rev. B {\bf 56}, 877 (1997).
\bibitem{ZX95} A. G. Loeser, {\em et al.}, Science, {\bf 273}, 325 (1996);
H. Ding, {\em et al.}, Nature, {\bf 382}, 51 (1996).
\bibitem{MBP93}
P. Monthoux, A. Balatsky, and D. Pines, Phys. Rev. Lett. {\bf 67}, 3448 (1993);
Phys. Rev. B {\bf 46}, 14803 (1992).
\bibitem{MP93} 
 P. Monthoux and D. Pines, Phys. Rev. B {\bf 47}, 
 6069 (1993); {\em ibid} {\bf 48}, 4261 (1994).
\bibitem{MMP90} A. Millis, H. Monien, and D. Pines, Phys.
 Rev. B {\bf 42}, 1671 (1990).
\bibitem{SP96} B. P. Stojkovi\'c and D. Pines, Phys. Rev. Lett.
 {\bf 76}, 811 (1996); Phys. Rev. B {\bf 55}, 8576 (1997).
\bibitem{CPS96}A. V. Chubukov, D. Pines, and B. P. Stojkovi\'c, 
J. Phys.: Condens. Matter {\bf 8}, 10017 (1996).
\bibitem{Sad79} M. V. Sadovskii, Sov. Phys. JETP {\bf 50}, 989 (1979);
J. Moscow Phys. Soc. {\bf 1}, 391 (1991);  see also
R. H. McKenzie and D. Scarratt, Phys Rev. B {\bf 54}, R12709 (1996).
\bibitem{comment} Details of the derivation of Eq.~\ref{sadself}
 will be given elsewhere.
\bibitem{CM97}  A. Chubukov, D. Morr, and K. A. Shakhnovich,
 Phil. Mag. B {\bf 74}, 563 (1994);
 A. Chubukov and D. Morr, Phys. Rep. ({\em in the press}).
\bibitem{Bob} For a discussion of the case of the symmetry broken state
see J.\ R.\ Schrieffer, J. Low Temp. Phys. {\bf 99}, 397 (1995).
\bibitem{MP94}P. Monthoux and  D. Pines, Phys. Rev. B  {\bf 50},
16015, (1994).
\bibitem{KS90} A. Kampf and  J. R. Schrieffer, Phys. Rev. B {\bf 42},
 7967 (1990).
\bibitem{Cub97}A. V. Chubukov, {\em private communication}.
 \end{thebibliography}
\end{document}